\documentclass[11pt]{article}

\usepackage{amsfonts}
\usepackage{amsbsy} 
\usepackage{epsfig}
\usepackage{latexsym}
\input amssym.def
\input amssym.tex

\def\bequ{\begin{equation}}
\def\eequ{\end{equation}}
\def\barr{\begin{array}}
\def\earr{\end{array}}

\def\ben{\begin{equation}}
\def\een{\end{equation}}
\def\bena{\begin{eqnarray}}
\def\eena{\end{eqnarray}}

\newcommand{\sect}[1]{\setcounter{equation}{0}\section{#1}}

\advance \topmargin by -\headheight
\advance \topmargin by -\headsep
\evensidemargin \oddsidemargin
\advance\hoffset by -3mm  
\advance\voffset by  8mm  

\def\b1{e^0}

\begin{document}
\hfuzz=100pt
\title{{\Large \bf{The Melvin Universe in Born-Infeld Theory \\ and other Theories of Non-Linear Electrodynamics}}} 
\author{G W Gibbons\footnote{Email address: gwg1@damtp.cam.ac.uk} \ \& C A R Herdeiro \footnote{Email address: car26@damtp.cam.ac.uk}
\\ (D.A.M.T.P.)
\\Department of Applied Mathematics and Theoretical Physics,
\\ Centre for Mathematical Sciences,
\\ University of Cambridge, 
\\ Wilberforce Road,
\\ Cambridge CB3 0WA,
 \\ U.K.}

\date{January 2001}

\maketitle
\centerline{DAMTP-2001-9}

\begin{abstract} 
We derive a Melvin Universe type solution describing a magnetic field permeating the whole Universe in gravity minimally coupled to \textit{any} non-linear electromagnetic theory, including Born-Infeld Theory. For a large set of non-linear electrodynamics theories, our solution is complete and  non-singular, as long as the magnetic field is sub-critical. We examine some properties of the solution; in particular there is a shift of the symmetry axis and a non-standard period along the orbits of the $U(1)$ symmetry to avoid a conical singularity. We show these are consistent with the usual Dirac quantization condition for the magnetic flux. We find exact solutions describing propagation of waves in the `generalized Melvin Universe' along the principal null directions of the electromagnetic field, where the Boillat and Einstein light-cone touch. By electric-magnetic duality we show that similar Melvin electric and dyonic universes can be obtained.
\end{abstract}


\sect{Introduction}

The `Melvin magnetic universe' is a simple solution to Einstein-Maxwell theory (or the bosonic sector of N=2 Supergravity) describing a bundle of magnetic flux lines in gravitational-magnetostatic equilibrium \cite{mel}\footnote{See also comments in \cite{mel2} referring to work previous to Melvin's on the same solution.}. Subsequent work has generalized the solution in different directions, including rotating and time dependent magnetic universes \cite{melgar}, black hole spacetimes which are asymptotically Melvin \cite{ernst}, gravitational waves travelling in the magnetic universe \cite{waves} and a Melvin universe with a non-trivial axion and dilaton in heterotic string theory \cite{tseyt}. Particularly interesting is the Melvin Universe solution in Kaluza-Klein theory since it derives from a flat spacetime by performing $U(1)$ Kaluza-Klein reduction with a twist in the identifications \cite{dowker}. Recently \cite{migo} a version of the Kaluza-Klein Melvin Universe was derived from M-theory as a solution to either type IIA or type 0A string theory.

One of the most interesting applications of the Melvin Universe is the study of black hole pair creation in a magnetic field. Although this spacetime configuration is stable against arbitrarily large radial perturbations \cite{thorne}, it is unstable against the non-perturbative process of black hole pair creation. This process was first discussed in \cite{karpacz}, based on the `asymptotically' Melvin `C-metric' type solution found by Ernst \cite{Cmelvin}. The Euclideanized version of this metric is in fact the instanton describing the creation in the Melvin Universe of a pair of  magnetically charged extreme Reissner-Nordstr\"om black holes with opposite charges \cite{strom}. In the Kaluza-Klein cases \cite{decay}, a second decay mode is by bubble nucleation. Moreover, and quite strikingly, the Ernst instanton is nothing else than the Kerr solution in the higher dimension \cite{decay}. In M-theory the instability produces pairs of D6-branes and anti-branes in the type IIA case, whereas for the type 0A the spacetime decays via bubble nucleation.

In this note we generalise the Melvin spacetime in yet a different direction - as a solution to gravity minimally coupled to any theory of non-linear electrodynamics (NLE) in four spacetime dimensions. Our approach relies on a limiting procedure in Einstein-Maxwell theory which transforms the Reissner-Nordstr\"{o}m solution into the Melvin solution. Since the Reissner-Nordstr\"om solution is known for an arbitrary theory of non-linear electrodynamics (see \cite{GR} which includes the original references), it is straightforward to generalise the procedure to Einstein-NLE and hence find the generalized Melvin solution.

Our motivation stems both from the fact that it might help us understand phenomena like black hole pair creation in gravity coupled to NLE, most notably Born-Infeld theory, and in connection to our recent work on causality in string theory and NLE \cite{gibher}. Therein we showed that the open string metric that has recently arisen in the work of Seiberg and Witten \cite{seiwit} is just the Boillat metric \cite{boillat} describing the propagation of fluctuation of the electromagnetic field in a non-trivial background electromagnetic field in NLE. The Boillat metric is not, in general, conformal to the Einstein or closed string metric. Hence, we expect light and gravitons to travel at different speeds, except along the principal null directions of the background electromagnetic field \cite{gibher}. The generalized Melvin solution we obtain in this paper will allow us to give examples of gravitational waves travelling in a curved spacetime along the direction in which the Boillat and Einstein light-cone coincide.

This paper is organised as follows. In section 2 we describe the formalism of non-linear electrodynamics coupled to gravity and the spherically symmetric charged solution which will be essential for our derivation. In section 3 we describe the limiting procedure and derive the solution. In section 4 we discuss some properties of the solution and compare it to the Maxwell case. In particular, we discuss the existence of curvature singularities beyond the critical field in Born-Infeld theory. A general shift of the symmetry axis is found. In order to avoid conical singularities a non-standard period must be imposed along the orbits of the azimuthal Killing vector field. We also discuss quantization of magnetic flux, showing that the standard condition is still obtained. In section 5 we find exact solutions describing gravitational waves in the generalized Melvin Universe. In section 6 we use electro-magnetic duality to derive electric and dyonic Melvin Universes. We close with a discussion.

\sect{General Relativity coupled to non-linear electrodynamics}
\subsection{General Framework}
If we consider gravity minimally coupled to non-linear electrodynamics, the action is
\bequ
S=\frac{1}{16\pi G_N}\int d^4x \sqrt{-g}R+\frac{1}{4\pi}\int d^4x \sqrt{-g}L(x,y).
\label{gpnle}
\eequ
The quantities $x\equiv\frac{1}{4}F_{\mu \nu}F^{\mu \nu}$ and $y\equiv \frac{1}{4}F_{\mu \nu}\star F^{\mu \nu}$ are the two only independent relativistic invariants constructed from the Maxwell field in four spacetime dimensions, and $L(x,y)$ an arbitrary function of those variables. We will take Newton's constant $G_N=1$.

The equations of motion derived from this action are more easily written in terms of the Legendre dual description of non-linear electrodynamics (see, for instance, \cite{gibher}). This involves introducing the tensor $P_{\mu \nu}$ defined by
\bequ
dL=-\frac{1}{2}P^{\mu \nu}dF_{\mu \nu}.
\label{dualm}
\een
$P_{\mu \nu}$ coincides with $F_{\mu \nu}$ for Maxwell's theory. In general it reads
\ben
P_{\mu \nu}= - \bigl ( L_x F_{\mu \nu} + L_y \star F_{\mu \nu} \bigr ),
\label{constitutive}
\een
where subscripts on $L$ denote differentiation. The components of $P_{\mu\nu}$ are just $\bf D$ and $\bf H$, the electric induction and magnetic field respectively and hence (\ref{constitutive}) are faced as the constitutive relations for the non-linear electromagnetic theory. We then denote by $s$ and $t$ the two independent invariants in terms of the dual variables, i.e., $s\equiv-\frac{1}{4}P_{\mu \nu}P^{\mu \nu}$ and $t\equiv-\frac{1}{4}P_{\mu \nu}\star P^{\mu \nu}$. The covariant Hamiltonian $\hat{L}$,
\bequ
\hat{L}=-\frac{1}{2}P^{\mu \nu}F_{\mu \nu}-L,
\eequ
is a function of $(s,t)$. In an orthonormal frame, $\hat{L}$ coincides with the usual Hamiltonian for purely electric configurations and minus the usual Hamiltonian for purely magnetic ones. The equations of motion for the coupled system are \cite{plebanski}
\bequ
D_{\mu}P^{\mu \nu}=0, \ \ \ \ \ \ \ \ R_{\mu \nu}-\frac{1}{2}Rg_{\mu \nu}=8\pi T_{\mu \nu},
\label{equmot}
\eequ
with the energy momentum tensor\footnote{We use a mostly plus signature. The Maxwell Lagrangian is $L=-x=s$.}
\bequ
T_{\mu \nu}= \frac{1}{4\pi}\left(\hat{L}_s P_{\mu \alpha} g^{\alpha \beta} P_{\nu \beta}+g_{\mu \nu}\left( 2s\hat{L}_s+t\hat{L}_t-\hat{L} \right)\right).
\label{emtensor}
\eequ

\subsection{Spherically symmetric solution}
The spherically symmetric solution has been known since the 1930's. One finds that Birkhoff's theorem holds. For a general theory, the electrically charged solution  may be cast in the form (see \cite{GR} which includes the original references)
\ben
\barr{c}
\displaystyle{ds^2_{\rm Einstein}= -\left(1 -{2 m(r)\over r}\right) dt^2 +
 \left(1 -{2 m(r) \over r}\right)^{-1}dr^2 +r^2  (d \theta ^2 + \sin ^2 \theta d \phi^2 )}, \\\\ 
\displaystyle{P= {Q dt\wedge d r  \over { r^2}}}. 
\label{rnnle}
\earr
\een
The latter is the radial component of the electric induction ${\bf D}$. The only non-trivial invariant is then 
\bequ
s={Q^2 / (2 r^4)}.
\label{inv}
\eequ
The metric function that defines the solution obeys the differential equation
\ben
\frac{dm(r)}{dr}=  r^2 \hat{L}(s),
\label{einszz}
\een
for $s$ as in (\ref{inv}). These equations permit one to read off the behaviour of the solutions
given the function ${\hat L}(s)$. But it is actually easier to implement another program: choose $m(r)$ such that the geometry is non-singular and find the `theory' that gives rise to such non-singular solution. This program has been implemented in some recent papers \cite{beato}. Therein, possibilities for $m(r)$ are considered so as to approach the usual Reissner-N\"ordstrom solution at large $r$. Furthermore, the geometry is chosen as to yield no curvature singularities; all scalar polynomials in the curvature are bounded for small $r$. One then finds the theory for which such geometries are solution via (\ref{einszz}). Although one may construct interesting solutions in this fashion, resembling Bardeen tunnels, they derive from obscure theories of non-linear electrodynamics.

\sect{The Melvin Universe}

We now derive the generalized Melvin Universe solution to non-linear electrodynamics coupled to gravity. We start by describing a limiting procedure that generates the Melvin Universe of Einstein-Maxwell theory from the Reissner-N\"ordstrom solution. Then we apply the same procedure to the non-linear electrodynamics case, and verify the solution by solving the Einstein equations. The fields we obtain take the form:

\bequ
\barr{c}
\displaystyle{ds^{2}=\frac{\rho^2+\frac{4}{H^2}h(\rho)}{\Lambda(\rho)^2}d\phi^2+\frac{\rho^2\Lambda(\rho)^2}{\rho^2+\frac{4}{H^2}h(\rho)}d\rho^2+\Lambda(\rho)^2\left(-dT^2+dZ^2\right)},
\\\\ 
\displaystyle{P=-\frac{H\rho}{\Lambda(\rho)^2}d\phi \wedge d\rho},
\label{genmel}
\earr
\eequ
where the functions $h(\rho)$ and $\Lambda(\rho)$ are given by
\bequ
h(\rho)=\left[1-\frac{s^{-\frac{1}{4}}}{4}\int_0^s \hat{L}(s')s'^{-\frac{7}{4}}ds'\right]_{s=-\frac{H^2}{2\Lambda(\rho)^4}}, \ \ \ \ \ \Lambda(\rho)=1+\frac{H^2}{4}\rho^2.
\label{funct}
\eequ
For Maxwell's theory of electromagnetism, $\hat{L}=s$ and so one may easily check that $h(\rho)=0$. The expression for $h(\rho)$ might seem ill-defined, since it involves quartic roots that should be evaluated at a negative quantity. It should be understood, however, that first one should evaluate the integral and only then replace $s$ by its value. In fact the expression is just a restatement of the infinite sum (\ref{hrho}). The constant $H$ in the solution defines the magnetic field strength at the origin, which is, in general, no longer at $\rho=0$ as shown below.

\subsection{Melvin Universe in Einstein-Maxwell theory}
Let us describe the limiting procedure that generates the Melvin universe starting with the Reissner-N\"ordstrom (RN) solution. Start by writing the RN solution using stereographic coordinates for the $S^2$ sections:
\bequ
\barr{c}
\displaystyle{ds^2=-Vdt^2+V^{-1}dr^2+\frac{4r^2}{(1+\zeta \bar{\zeta})^2}d\zeta d\bar{\zeta},
\ \ \ \ \ \ V\equiv1-\frac{2M}{r}+\frac{Q^2}{r^2}},
\\\\ \displaystyle{F=-\frac{Q}{r^2}dr\wedge dt},
\label{usualrn}
\earr
\eequ
Perform the diffeomorphism given by 
\bequ
\tilde{\zeta}=\frac{\zeta}{\lambda}, \ \ \ \ \ \tilde{r}=r\lambda, \ \ \ \ \ \tilde{t}=\frac{t}{\lambda},
\label{fct}
\eequ
to get the RN solution in $(\tilde{t}, \tilde{r}, \tilde{\zeta}, \bar{\tilde{\zeta}})$ coordinates. Next we want to take the following limit:
\bequ
\lambda \rightarrow 0, \ \ \ \ \ M,Q \rightarrow \infty, \ \ \ \ \ M\lambda^3=fixed\equiv m, \ \ \ \ \ Q\lambda^2=fixed\equiv q.
\label{limit}
\eequ
That the resulting spacetime configuration remains a solution requires proof. We will come back to this point below. Notice that in the limit the $S^2$ is `flattened' into $~{\Bbb R}^2$. This changes the symmetry from spherical to axial. Denote $\tilde{\zeta}=x+iz$  and perform the final coordinate transformations and relabellings:
\bequ
\barr{c}
\displaystyle{\tilde{t}=i\phi\frac{2}{B^2}, \ \ \ \ x=i\frac{T}{2}, \ \ \ \ \ z=\frac{Z}{2}, \ \ \ \ \ \tilde{r}=1+\frac{B^2}{4}\rho^2\equiv \Lambda(\rho)},
\\\\  \displaystyle{q=iB, \ \ \ \ \ m=-B^2/2}. 
\label{sct}
\earr
\eequ
Notice the change from timelike to spacelike character of $\tilde{t}$. This will change the electromagnetic field from electric to magnetic. The resulting configuration is the Melvin Universe \cite{mel}:
\bequ
\barr{c}
\displaystyle{ds^2=\rho^2\Lambda(\rho)^{-2}d\phi^2+\Lambda(\rho)^2\left[-dT^2+dZ^2+d\rho^2\right]},
\\\\ F=B\rho\Lambda(\rho)^{-2} d\rho\wedge d\phi.
\label{emmel}
\earr
\eequ
The constant $B$ defining the solution is the magnetic induction strength at the origin, since ${\bf B}=B\Lambda(\rho)^{-2}{\bf \partial_z}$.

\subsection{Generalized Melvin Universe}

\subsubsection{The Limiting Procedure}
Now consider gravity coupled to general non-linear electrodynamics as described by (\ref{gpnle}). The generalized RN solution is (\ref{rnnle}). It takes the form (\ref{usualrn}) with 
\bequ
V=1-\frac{2M}{r}+\frac{I(r^{-1})}{r^2},
\eequ
and $F\rightarrow P$ the dual Maxwell tensor. Equation (\ref{einszz}) translates into 
\bequ
r^{-2}\frac{dI(r^{-1})}{d(r^{-1})}+r^{-1}I(r^{-1})=2r^3\hat{L}\left(\frac{Q^2}{2r^4}\right).
\label{difeq}
\eequ
for the function $I(r^{-1})$. Next perform (\ref{fct}). In order to solve the differential equation we expand $I(\lambda \tilde{r}^{-1})$ in a power series (around spatial infinity) and $\hat{L}(s)$ in its Taylor series around zero  (we are assuming $\hat{L}(s)$ to be analytic; we will be able to relax this assumption in the final result):
\bequ
I(\lambda \tilde{r}^{-1})=Q^2\sum_{n=0}^{\infty}f_n (\lambda \tilde{r}^{-1})^n, \ \ \ \ \ \hat{L}(s)=\sum_{m=1}^{\infty}\frac{s^m}{m!}\left[\frac{\partial^{m} \hat{L}(s)}{\partial s^{m}}\right]_{s=0}.
\eequ
Requiring that $\hat{L}$ approach the value for Maxwell's theory in the weak field limit yields $[\partial \hat{L} / \partial s]_{s=0}=1$ and for the usual RN solution to hold one must have $f_0=1$. Substituting in (\ref{difeq}) and matching powers of $\lambda r^{-1}$ we find
\bequ
I(\lambda \tilde{r}^{-1})=Q^2\sum^{\infty}_{n=0}\frac{1}{(1+n)!(1+4n)}\left(\frac{Q^2\lambda^4}{2\tilde{r}^4}\right)^n\left[\frac{\partial^{n+1} \hat{L}(s)}{\partial s^{n+1}}\right]_{s=0}.
\eequ
Taking the limit (\ref{limit}), which for the function $I$ yields
\bequ
\lim_{\lambda \rightarrow 0, Q \rightarrow \infty, \lambda^2 Q=q} \lambda^4 I(\lambda \tilde{r}^{-1})=q^2\sum^{\infty}_{n=0}\frac{1}{(1+n)!(1+4n)}\left(\frac{q^2}{2\tilde{r}^4}\right)^n\left[\frac{\partial^{n+1} \hat{L}(s)}{\partial s^{n+1}}\right]_{s=0}.
\label{limi}
\eequ
Finally perform (\ref{sct}) with relabelling $B\rightarrow H$ to obtain the generalized Melvin magnetic universe (\ref{genmel}). The function $h(\rho)$ is 
\bequ
h(\rho)=-\sum^{\infty}_{n=1}\frac{1}{(1+n)!(1+4n)}\left(-\frac{H^2}{2\Lambda(\rho)^4}\right)^n\left[\frac{\partial^{n+1} \hat{L}(s)}{\partial s^{n+1}}\right]_{s=0}.
\label{hrho}
\eequ
It can also be expressed in the more useful integral form (\ref{funct}).

\subsubsection{Solving the Einstein equations}
As we mentioned above, it is not clear that after performing the limit (\ref{limit}) one ends up with a solution of the equations of motion. The Maxwell equation in (\ref{equmot}) is clearly obeyed by (\ref{genmel}) since it is completely identical to the usual Melvin solution and the determinant of the generalized Melvin metric is unchanged from the usual Melvin solution. So we just need to show the Einstein equations are solved. The Einstein tensor for the geometry (\ref{genmel}) with $\Lambda(\rho)$ given by (\ref{funct}) but unspecified $h(\rho)$ is diagonal and so is the energy-momentum tensor. The $\rho \rho$ and $\phi \phi$ equations of motion are equivalent, yielding the first order ordinary differential equation (O.D.E.)
\bequ
4s^2\frac{d h(s)}{ds}-s+s h(s)=-\hat{L}(s).
\label{ode1}
\eequ
We are using the form of the only non-vanishing invariant
\bequ
s=-\frac{H^2}{2\Lambda(\rho)^4},
\eequ
to express $h(\rho)$ as a function of $s$. It is easy to verify that $h(\rho)$ given by (\ref{funct}) (without specifying the value of $s$ in terms of $\rho$) is the solution. Notice that this equation does not define the lower limit in the integral (\ref{funct}); this is set by the requirement that one should recover the usual Melvin spacetime when $\hat{L}$ to correspond to Maxwell's theory. We can therefore drop the analyticity assumption for $\hat{L}(s)$. The $tt$ and $zz$ equations are also equivalent, yielding the second order O.D.E.
\bequ
-8s^3\frac{d^2 h(s)}{ds^2}-14s^2\frac{d h(s)}{ds}+s(1-h)=2s\hat{L}_s -\hat{L}.
\eequ
This equation is not independent of the first order O.D.E. (\ref{ode1}). Using the latter to find an expression for $\hat{L}_s$ and replacing this in the second order O.D.E. one recovers the first order equation. This is similar to what happens for solution (\ref{rnnle}). Equation (\ref{einszz}) comes from the $tt$ or $rr$ components of the Einstein equations, whereas the $\theta \theta$ and $\phi \phi$ components give a second order equation which is obtained by differentiating (\ref{einszz}). Hence we have just shown that indeed the fields (\ref{genmel}) are a solution to the equations (\ref{equmot}).

\sect{Properties of the solution and special cases}
We now discuss some properties of the solution (\ref{genmel}). We will enphasize two particular types of theories of non-linear electrodynamics:
\begin{description} 
\item[i)] The family described by the Lagrangian:
\bequ
L=-x+\frac{a}{2}x^2+\frac{b}{2}y^2,
\eequ
which has a dual Lagrangian (up to terms of higher order in the constants $a,b$),
\bequ
\hat{L}=s-\frac{a}{2}s^2-\frac{b}{2}t^2.
\eequ
The Euler-Heisenberg theory (see for instance \cite{zuber}) is the special case with $a=8\alpha^2/(45m^4)$ and $b=14\alpha^2/(45m^4)$, where $m$ is the electron mass and $\alpha$ the fine structure constant. The function $h(\rho)$ in (\ref{funct}) becomes
\bequ
h(\rho)=-\frac{aH^2}{20\Lambda(\rho)^4}.
\label{heh}
\eequ
\item[ii)] The Born-Infeld Lagrangian:
\bequ
L=\frac{1}{\beta^2}\left(1-\sqrt{1+2\beta^2x-\beta^4y^2}\right),
\eequ
which has an \textit{exact} dual Lagrangian of the same form
\bequ
\hat{L}=\frac{1}{\beta^2}\left(\sqrt{1+2\beta^2s-\beta^4t^2}-1\right).
\eequ
In string theory $\beta=2\pi\alpha'$. The integral (\ref{funct}) does not have a closed form and the best one can do in this case is to re-express $h(\rho)$ in terms of the infinite sum (\ref{hrho}):
\bequ
h(\rho)=-\sum^{\infty}_{n=1}\frac{1}{(1+n)!(1+4n)}\left(\frac{H^2\beta^2}{2\Lambda(\rho)^4}\right)^n\left[\prod_{m=1}^{n}(2m-1)\right].
\label{hbi}
\eequ

\end{description}
Notice that both (\ref{heh}) for the Euler-Heisenberg case and (\ref{hbi}) are negative for all values of $\rho$.

\subsection{Singularities and asymptotics}
The Einstein-Maxwell Melvin Universe (\ref{emmel}) represents a parallel bundle of magnetic flux held together by its own gravitational attraction. It is a totally non-singular configuration. It is easy to check the absence of singularities in scalar polynomials of the curvature:
\bequ
R=0, \ \ R_{\mu \nu}R^{\mu \nu}=4B^4\Lambda(\rho)^{-8}, \ \  R_{\mu \nu \alpha \beta}R^{\mu \nu \alpha \beta}=4B^4\Lambda(\rho)^{-8}(3B^4\rho^4-24B^2\rho^2+80).
\eequ
That $R=0$ follows from the conformal invariance of Maxwell's theory in four spacetime dimensions. The non-singular nature of the solution motivated Melvin to dub his solution a magnetic `geon'. 

For the generalised Melvin solution, the existence of singularities will obviously depend on the choice of Lagrangian. From the Einstein equations (\ref{equmot}) one reads for the Ricci scalar
\bequ
R=8\left[\hat{L}(s)-s\frac{d\hat{L}(s)}{ds}\right]_{s=-\frac{H^2}{2\Lambda(\rho)^4}},
\eequ
so that if the energy density or its gradient blows up so does the curvature. For the Euler-Heisenberg family this curvature invariant is bounded everywhere, but for the Born-Infeld case it diverges at radial distance
\bequ
\Lambda(\rho)^4=\beta^2 H^2.
\eequ
Therefore, if the constant $H$ in the solution is such that $H^2>\beta^{-2}$, then the solution is singular. The value $H^2=\beta^{-2}$ is the maximal magnetic field for a purely magnetic configuration in Born-Infeld theory. Notice that Born-Infeld theory has a maximal value for the electric field $\vec{E}$ and for the magnetic field $\vec{H}$ (but not for the inductions). Thus, the free parameter describing the solution is constrained if we require absence of curvature singularities. In general, we expect that if the non-linear theory of electrodynamics under consideration has a critical field the curvature will diverge at that critical field.

Independently of the existence of curvature singularities, the metric (\ref{genmel}) will be singular if there exists a radial distance $\rho_0$ such that 
\bequ
\rho_0^2+\frac{4}{H^2}h(\rho_0)=0.
\label{coosing}
\eequ
This is certainly the case for Euler-Heisenberg and Born-Infeld, since $h(\rho)$ is negative everywhere for these two theories. Consider the $\rho$-$\phi$ sections of the metric (\ref{genmel})
\bequ
ds^2=\frac{\rho^2}{g_{\phi \phi}}d\rho^2+g_{\phi \phi}d\phi^2.
\eequ
In the neighbourhood of $\rho=\rho_0$,
\bequ
ds^2\simeq \frac{\rho_0^2}{\dot{g}_{\phi \phi}(\rho-\rho_0)}d\rho^2+\dot{g}_{\phi \phi}(\rho-\rho_0)d\phi^2, 
\eequ
where $\dot{g}_{\phi \phi}=(dg_{\phi \phi}/d\rho)$ at $\rho=\rho_0$, which we assume to be non-zero. Now introduce the new coordinate $R$ by
\bequ
(\rho-\rho_0)=4\dot{g}_{\phi \phi}\frac{R^2}{\rho_0^2},
\eequ
to get the metric 
\bequ
ds^2\simeq dR^2+R^2d\left(\frac{2\dot{g_{\phi \phi}}}{\rho_0}\phi\right).
\eequ
Hence, if we identify $\phi$ with period 
\bequ
\delta \phi=\frac{\pi \rho_0}{\dot{g}_{\phi \phi}},
\eequ
the singularity at $\rho=\rho_0$ is just the usual singularity at the origin of flat space in polar coordinates. Any other period for $\phi$ leads to a conical singularity. In any case, the point $\rho=\rho_0$ should be faced as the location of the axis of symmetry, where $\rho_0$ is the first solution of (\ref{coosing}) when coming from infinity. The region $0<\rho<\rho_0$ will be discussed below.

Asymptotically, our generalized Melvin Universe will approach the usual Melvin solution, which is clearly a consequence of the requirement that the theory of non-linear electrodynamics under consideration should approach Maxwell's theory for weak enough fields. The components of the Riemann tensor in an orthonormal frame vanish as $1/\rho^6$. In this sense, asymptotic Melvin spacetimes are asymptotically flat (away from the axis of symmetry). The components of the Weyl tensor in an orthonormal frame, $C^{a}_{\ \ b cd}$, also vanish with the same power of $\rho$. The fact that this is true for all non-vanishing components indicates that there might exists a conformal compactification of the spacetime, which however is not know. If it exists the conformal spacetime should have a light-like conformal boundary.

We also note that the isometry group of the solution (\ref{genmel}) is the same as the one of (\ref{emmel}). There is a  $U(1)\times ISO(1,1)$ isometry corresponding to a circle action along the $\phi$ direction, boosts along the t-z direction and translations along the t and z directions.

\subsection{Test motions and supersymmetry}

In the Newtonian limit, the generalised Melvin Universe has a gravitational potential equal to the usual Melvin solution, which is of the form
\bequ
V=\frac{H^2}{4}\rho^2+\frac{H^4}{32}\rho^4.
\eequ
This potential well will tend to confine the particles in a cylinder of finite radius. In this respect the behaviour is similar to that of Anti-de-Sitter spacetime, despite the fact that AdS has constant curvature in contrast to the asymptotically vanishing curvature of Melvin discussed in the previous subsection. For the full relativistic analysis consider the geodesic equations, where $\lambda$ is an affine parameter along the trajectory:
\bequ
\barr{c}
\displaystyle{\left(\frac{d\rho}{d\lambda}\right)^2=\frac{\rho^2+4H^{-2}h(\rho)}{\rho^2\Lambda(\rho)^2}\left[\frac{1}{\Lambda(\rho)^2}(E^2-p^2)-m^2\right]-\frac{l^2}{\rho^2}},
\\\\
\displaystyle{\frac{d\phi}{d\lambda}=\frac{l\Lambda(\rho)^2}{\rho^2+4H^{-2}h(\rho)}, \ \ \  \frac{dt}{d\lambda}=\frac{E}{\Lambda(\rho)^2}, \ \ \ \ \frac{dz}{d\lambda}=\frac{p}{\Lambda(\rho)^2}}.
\label{geode}
\earr
\eequ
The conserved quantities $E,p,l$ are associated with the Killing vector fields $\partial / \partial t, \partial / \partial z$ and $\partial / \partial \phi$ respectively. The `mass' $m$ is as usual associated with the metric Killing tensor. The $t,z$ coordinates are the proper time and proper radial distance for an observer at the origin. As seen by this observer clocks tick faster as the radial distance increases. This is exactly the opposite of the redshift witnessed by the observer at infinity for events close to the black hole horizon in a Schwarzchild spacetime.

From the radial equation we can see that at large distances the repulsive terms dominate over the attractive (energy) term. Hence, no particles may enter the large enough $\rho$ region, whatever values the particle's quantum numbers take. The solely exception is a photon with no angular momentum not moving entirely in the $z$ direction, which is not confined to a cylinder of some radius. In this way we see that the Newtonian intuition is mostly confirmed. This resembles the Melvin case \cite{melgeo} due to the similar asymptotics. 

The behaviour for small $\rho$ is quite distinct from the usual Melvin solution. From the radial equation we can see that no particle can cross the $\rho=\rho_0$ cylinder. But geodesics may exist on both sides of this cylinder. As an example consider the Euler-Heisenberg case. Equation (\ref{coosing}) only has one solution. Its left hand side is negative in $0<\rho<\rho_0$ and positive for $\rho>\rho_0$. Bound states with $E^2<m^2$ can therefore exist within the $\rho=\rho_0$ cylinder. In a sense this is analogous to the $r<0$ region in the Kerr metric. Although  $\rho=\rho_0$ should be seen as the location of the axis of symmetry (like the $r=0$ in Kerr), there is a region beyond it for which the metric is different and therefore taking it seriously is not double counting (like the negative $\rho$ region for flat space in polar coordinates). Hence the spacetime can have two geodesically complete disjoint regions.

To close this section we remark that on a curious effect for the Einstein-Maxwell Melvin Universe: at any radial distance one may find a `no-force' configuration, i.e. a static test particle. The charged geodesics are
\bequ
 \barr{c}
\displaystyle{\left(\frac{d\rho}{d\lambda}\right)^2=\frac{1}{\Lambda(\rho)^4}(E^2-p^2)-\frac{m^2}{\Lambda(\rho)^2}-\frac{1}{\rho^2}\left(l-\frac{2q}{B\Lambda(\rho)}\right)^2},
\\\\
\displaystyle{\frac{d\phi}{d\lambda}=\frac{\Lambda(\rho)^2}{\rho^2}\left(l-\frac{2q}{B\Lambda(\rho)}\right)},
\label{chageo}
\earr
\eequ
where $q$ is the particle's charge. The $t$ and $z$ equations are the same as in (\ref{geode}). Notice that the charged geodesics for the generalised Melvin Universe are more subtle due to the coupling to the Maxwell potential. Choosing $l,q$ as to cancel the right hand side of the $\phi$ equation of motion (\ref{chageo}) for a given $\rho$, putting $p=0$ to cancel $z$ motion (\ref{geode}) and choosing $E^2=\Lambda(\rho)^2 m^2$, the particle becomes static. Obviously this cannot happen for photons ($m=0$). 

The possibility of a no-force configuration, together with the fact that the Reissner-N\"ordstrom family includes a supersymmetric solution raises a question about the supersymmetry of the Melvin solution within $N=2$, $D=4$ supergravity. The Killing spinor equation is, in the notation of forms,
\bequ
\left(d+\frac{1}{4}\omega_{ab}\Gamma^{ab}-\frac{1}{4}F_{ab}\Gamma^{ab} \Gamma\right)\epsilon=0.
\eequ
We use for the flat metric $\eta_{ab}=diag(-1,1,\rho^2,1)$, i.e., flat space in cylindrical coordinates. Then, one easily checks that the Killing spinor equation implies the conditions:
\bequ
\epsilon=\Lambda^{\frac{1}{2}}\epsilon_0, \ \ \ \ \left({\bf 1}-\frac{2}{B}\Gamma^{\phi}\right)\epsilon_0=0,
\label{kilspi}
\eequ
where $\epsilon_0$ is a constant spinor. Keeping in mind that $\{ \Gamma^{\phi}, \Gamma^{\phi}\}=2\rho^{-2}$ one concludes that the right hand side equation has no solution all over the spacetime. 
However for any value of $B$ (\ref{kilspi}) is obeyed on the hypersurface $B^2\rho^2/4=1$. This hypersurface `preserves half of the vacuum supersymmetries'. Note that if we require the classical expression for circular orbits of a charged particle in a constant magnetic field, $\rho^2=l/(q\tilde{B})$, where $\tilde{B}$ is the magnetic field measured at radial coordinate $\rho$, the no-force condition $l=2q/(B\Lambda(\rho))$ singles out this `supersymmetric' hypersurface. In a subtle way, therefore, the no-force condition is still associated with unbroken supersymmetry of the solution. The phenomenon uncovered here resembles what are called hypersurface twistors \cite{tod}.

\subsection{Dirac Quantization of the Magnetic Flux}
Inverting the constitutive relations (\ref{constitutive}), one finds
\bequ
F_{\mu \nu}=\hat{L}_s P_{\mu \nu}+\hat{L}_t\star P_{\mu \nu}.
\eequ
We will assume that $\hat{L}_t=0$ for a purely magnetic configuration, which is true for most theories of interest (including Born-Infeld and Euler-Heisenberg). Then 
\bequ
F_{\mu \nu}=\hat{L}_s P_{\mu \nu} \ \ \ \Leftarrow \ \ \ A_{\infty}=-\frac{2}{H\Lambda(\rho)}\left(1-h(\rho)+\frac{\Lambda(\rho)}{2\rho H^2}\frac{dh}{d\rho}\right)d\phi,
\eequ
where $F=dA$. The subscript $\infty$ indicates this is a good gauge at infinity. This is not the case at $\rho_0$ as defined by (\ref{coosing}). A good gauge at $\rho_0$ is $A_0=A_{\infty}+d\chi$ with 
\bequ
\chi=\frac{2}{H}\left(1+\frac{1}{2\rho_0 H^2}\frac{dh}{d \rho}(\rho_0)\right)\phi.
\eequ
For fermions minimally coupled to the Maxwell field, this gauge transformation induces a change of phase in the wave function required by gauge invariance, of the form $\Psi(x)\rightarrow \exp{(ie\chi)}\Psi(x)$, where $e$ is the electron charge. To be singled valued we have the following quantization condition for the constant $H$
\bequ
\frac{ \delta \phi}{2\pi H}\left(1+\frac{1}{2\rho_0 H^2}\frac{dh}{d \rho}(\rho_0)\right)=\frac{n}{2e}.
\label{quantization}
\eequ
The left hand side of this equation turns out to be  the magnetic flux (of the magnetic induction $\vec{B}$) through a $z=const, t=const$ surface,
\bequ
\Phi_B=\frac{1}{8\pi}\int F_{\mu \nu}dS^{\mu \nu}=\frac{1}{4\pi}\int_{\theta=0}^{\delta \phi}\int_{\rho=\rho_0}^{\infty}F_{\rho \phi}d\rho d\phi.
\eequ
Thus, the flux for the magnetic induction is quantized, and one reads off the usual Dirac quantization condition,
\bequ
\frac{e\Phi_B}{\hbar c}=\frac{n}{2}, 
\eequ
where we have reinserted fundamental constants. In the Einstein-Maxwell case, the analogous condition to (\ref{quantization}) takes the form $2e=Bn$, and one gets the universal quantization condition for the flux from the fact that $\Phi_B=B^{-1}$. In the generalised case, the metric parameter $H$ obeys the more involved constraint (\ref{quantization}) but the universal flux quantization still holds.

\sect{Travelling Waves in the Magnetic Universe}
A spacetime solution to Einstein-Maxwell theory describing waves of arbitrary profile travelling in a magnetic universe was found in \cite{waves}. The solution is described by the fields $(\tilde{g}_{\mu \nu}, F_{\mu \nu})$ and is obtained by using a Kerr-Schild ansatz:
\bequ
\tilde{g}_{\mu \nu}=g_{\mu \nu}+\Lambda(\rho)^{-2}\Psi(u,\rho,\phi)k_{\mu}k_{\nu}.
\label{kerrsch}
\eequ  
The fields $(g_{\mu \nu}, F_{\mu \nu})$ are the ones in the usual Melvin solution (\ref{emmel}), and the vector $k=k^{\mu}\partial_{\mu}=\partial / \partial v$ is null and Killing for the Melvin geometry. We have introduced light-cone coordinates $(u,v)=(z \pm t)/\sqrt{2}$. The Einstein-Maxwell equations are then still solved if the function $\Psi(u,\rho,\phi)$ is harmonic in the usual Melvin spacetime, i.e. $D_{\mu}D^{\mu}\Psi(u,\rho,\phi)=0$, where $D_{\mu}$ is the covariant derivative for $g_{\mu \nu}$. The harmonic condition does not constrain the $u$ dependence (hence the arbitrary profile described by an arbitrary function $f(u)$), whereas the $\phi$ and $\rho$ dependence might be solved by separating variables. The geometry obtained reads
\bequ
\tilde{g}_{\mu \nu}dx^{\mu}dx^{\nu}=\Lambda(\rho)^2\left(2dudv+d\rho^2 +f(u)\cos{m(\phi-\phi_0)}P\left(\rho\right)du^2\right)+\frac{\rho^2d\phi^2}{\Lambda(\rho)^2}.
\eequ
The function $P(\rho)$ obeys a second order ordinary differential equation which can be computed to arbitrary accuracy by an iteration method described in \cite{waves}, whereas $\phi_0$ and $m$ are constants. The waves are travelling along the $k$ direction which lies on the Einstein light cone for the Melvin Universe. 

It is straightforward to generalise this solution to the non-linear case. Using again the ansatz (\ref{kerrsch}) for the metric, where now $g_{\mu \nu}$ is given by (\ref{genmel}) one can check that $(\tilde{g}_{\mu \nu}, P_{\mu \nu})$ is a solution to (\ref{equmot}), where $P_{\mu \nu}$ is still given by (\ref{genmel}), as long as the function $\Psi(u,\rho,\phi)$ is harmonic in the geometry of the generalized Melvin Universe. By separating variables we find that $\Psi(u,\rho, \phi)=f(u)R(\rho)\cos{m(\phi-\phi_0)}$ where $R(\rho)$ obeys:
\bequ
\frac{\partial}{\partial \rho}\left(\frac{\rho^2+4H^{-2}h(\rho)}{\rho}\frac{\partial R(\rho)}{\partial \rho}\right)-\frac{m^2 \rho \Lambda(\rho)^4}{\rho^2+4H^{-2}h(\rho)}R(\rho)=0.
\eequ
For large $\rho$ this approaches the solution discussed in \cite{waves}. The geometry describing waves in the generalized Melvin Universe is then
\bequ
d\tilde{s}^2=\Lambda(\rho)^2\left(2dudv+f(u)\cos{m(\phi-\phi_0)}R\left(\rho\right)du^2\right)+\frac{\rho^2\Lambda(\rho)^2}{\rho^2+4H^{-2}h(\rho)}+\frac{\rho^2+4H^{-2}h(\rho)} {\Lambda(\rho)^2}d\phi^2.
\eequ

To conclude this section, we remark that the metric describing the propagation of fluctuations in non-linear electrodynamics, the Boillat metric, is not conformal to the Einstein metric \cite{gibher}. The fact that the waves in the generalized Melvin Universe travel along a null direction of the metric (\ref{genmel}) is because this is the principal null direction of $F_{\mu \nu}$, and therefore along this direction the Einstein and Boillat light cones coincide. This would not be the case anymore if one could find solutions to (\ref{equmot}) of the Kerr-Schild type, where $k$ would have non-zero $\phi$ or $\rho$ components.

\sect{Electric-Magnetic Duality}
\subsection{Einstein-Maxwell}
The limiting procedure we described in section 3 also works if we start from the dyonic Reissner-N\"ordstrom black hole. In the usual stereographic coordinates this solution is
\bequ
\barr{c}
\displaystyle{ds^2=-Vdt^2+V^{-1}dr^2+\frac{4r^2}{(1+\zeta \bar{\zeta})^2}d\zeta d\bar{\zeta},
\ \ \ \ \ \ V\equiv1-\frac{2M}{r}+\frac{Q^2+P^2}{r^2}},
\\\\ \displaystyle{F=\frac{Q}{r^2}dt\wedge dr-P\sin\theta d\theta\wedge d\phi=\frac{Q}{r^2}dt\wedge dr -2iP\frac{d\zeta\wedge d\bar{\zeta}}{(1+\zeta\bar{\zeta})^2}}.
\label{magnern}
\earr
\eequ
Performing the method of section 3.1, taking also the limit $P\rightarrow \infty$ with $P\lambda^2=fixed \equiv p$, and performing a final set of coordinate transformations
\bequ
\barr{c}
\displaystyle{\tilde{t}=i\phi\frac{2}{E^2+B^2}, \ \ \ \ x=i\frac{T}{2}, \ \ \ \ \ z=\frac{Z}{2}, \ \ \ \ \ \tilde{r}=1+\frac{E^2+B^2}{4}\rho^2\equiv \Lambda(\rho)},
\\\\  \displaystyle{q=iB, \ \ \ \ \ p=iE, \ \ \ \ \ m=-(E^2+B^2)/2}, 
\earr
\eequ
we then end up with the dyonic version of the Melvin Universe
\bequ
\barr{c}
\displaystyle{ds^2=\rho^2\Lambda(\rho)^{-2}d\phi^2+\Lambda(\rho)^2\left[-dT^2+dZ^2+d\rho^2\right]},
\\\\ F=EdT\wedge dZ+B\rho\Lambda(\rho)^{-2}d\rho \wedge d\phi.
\label{elemel}
\earr
\eequ
Of course (\ref{magnern}) and (\ref{elemel}) can be obtained by $SO(2)$ duality rotations of Einstein-Maxwell theory in four dimensions from  (\ref{usualrn}) and (\ref{emmel}) respectively. Therefore the limiting procedure comutes with electric-magnetic duality.

\subsection{Non-linear electrodynamics}
For theories obeying the requirement $F_{\mu \nu}\star F^{\mu \nu}=P_{\mu \nu}\star P^{\mu \nu}$, there is also a electro-magnetic duality rotation that leaves invariant the equations of motion and energy-momentum tensor \cite{GR}. One therefore obtains new solutions to (\ref{equmot}). Born-Infeld is one such theory. We can thus obtain the dyonic Melvin Universe for Born-Infeld:
\bequ
\barr{c}
\displaystyle{ds^{2}=\frac{\rho^2+\frac{4}{H^2+D^2}h(\rho)}{\Lambda(\rho)^2}d\phi^2+\frac{\rho^2\Lambda(\rho)^2}{\rho^2+\frac{4}{H^2+D^2}h(\rho)}d\rho^2+\Lambda(\rho)^2\left(-dT^2+dZ^2\right)},
\\\\ 
\displaystyle{P=-\frac{H\rho}{\Lambda(\rho)^2}d\phi \wedge d\rho}+D dT \wedge dZ,
\label{genmelbi}
\earr
\eequ
with
\bequ
h(\rho)=-\sum^{\infty}_{n=1}\frac{1}{(1+n)!(1+4n)}\left(\frac{(H^2+D^2)\beta^2}{2\Lambda(\rho)^4}\right)^n\left[\prod_{m=1}^{n}(2m-1)\right], \ \ \Lambda(\rho)=1+\frac{D^2+H^2}{4}\rho^2.
\eequ
For theories without duality invariance one can obtain the Melvin electric universe starting with the magnetically charged Reissner-N\"ordstrom solution in the theory (see \cite{GR}) and applying the limiting procedure.

\sect{Discussion}
The limiting procedure we have described is very simple both mathematically and conceptually. However, notice it changes completely the causal structure and  singularity structure of the solution: in the linear case, i.e. Maxwell's theory, we start with a singular solution with a null surface and we end up with a completely non-singular solution with an everywhere timelike Killing vector field. With the wisdom of hindsight the procedure looks quite natural; but it is far from obvious that a charged black hole and a magnetic flux tube are so simply related.

In our generalisation of the procedure to the case of non-linear electrodynamics, taking the limit commutes with solving the equations of motion. However, one should not take this for granted. For instance, applying this procedure to a Reissner-N\"ordstrom-Anti de Sitter (or de Sitter) black hole spacetime , one ends up with
\bequ
\barr{c}
\displaystyle{ds^2=\left(\frac{1}{\Lambda(\rho)^2}-\frac{4\Omega\Lambda(\rho)^2}{3B^2\rho^2}\right)\rho^2d\phi^2+\left(\frac{1}{\Lambda(\rho)^2}-\frac{4\Omega\Lambda(\rho)^2}{3B^2\rho^2}\right)^{-1}d\rho^2+\Lambda(\rho)^2\left[-dT^2+dZ^2\right]},
\\\\ F=B\rho\Lambda(\rho)^{-2}d\rho\wedge d\phi,
\earr
\eequ
where $\Omega$ is the cosmological constant. In this case the equations of motion are only solved if $B^2=1$. The procedure might, however, be useful in finding other solutions of the flux tube type. It is perhaps worth exploring applications to black p-branes of string theory.

\section *{Acknowledgments}
C.H. is supported by FCT (Portugal) through grant no. PRAXIS XXI/BD/13384/97.

\end{document}